\documentclass[oneside,a4paper,12pt]{article}
\usepackage{latexsym}
\usepackage{euscript}
\usepackage{epsfig}
 \large\normalsize
\newcommand{\be}{\begin{equation}}
\newcommand{\ee}{\end{equation}}

\begin{document}
\begin{flushright}
UT-STPD-1/01\\
SISSA/18/2001/EP\\
SUSX-TH/01-013
\end{flushright}
\vskip 0.3cm
\begin{center}
{\Large \bf Leptogenesis in Smooth Hybrid Inflation}
\vskip 0.35cm
{R. Jeannerot$^{1}$, S. Khalil$^{2,3}$ and G. Lazarides$^{4}$}
\\ \vspace*{0.2cm} \small{\textit{$^1$SISSA, via Beirut 4, 
34014 Trieste, Italy.}}\\ \vspace*{1mm}
\small{\textit{$^2$Center for Theoretical Physics, University 
of Sussex, Falmer, Brighton, BN1 9QH, UK.}} \\  
\vspace*{1mm} \small{\textit{$^3$Ain Shams University, 
Faculty of Science, Cairo 11566, Egypt.}} \\ \vspace*{1mm}
\small{\textit{$^4$Physics Division, School of Technology, 
Aristotle University of Thessaloniki,\\ Thessaloniki 540 06, 
Greece.}}\\ \vspace*{0.5cm}

\begin{abstract}
{We present a concrete supersymmetric grand unified model based 
on the Pati-Salam gauge group $SU(4)_c \times SU(2)_L \times 
SU(2)_R$ and leading naturally to smooth hybrid inflation, which 
avoids the cosmological disaster encountered in the standard 
hybrid inflationary scenario from the overproduction of 
monopoles at the end of inflation. Successful `reheating' which 
satisfies the gravitino constraint takes place after the 
termination of inflation. Also, adequate baryogenesis via a 
primordial leptogenesis occurs consistently with the solar and 
atmospheric neutrino oscillation data as well as the $SU(4)_c$ 
symmetry.}
\end{abstract}

\setcounter{page}{1}
\end{center}

\thispagestyle{empty}

\par
The most promising theory for solving the cosmological magnetic 
monopole problem of grand unified theories (GUTs) is inflation. 
This theory also solves many of the outstanding problems of the 
standard Big Bang cosmological model and predicts the formation 
of the large scale structure in the universe as well as the 
temperature fluctuations which are observed in the cosmic 
microwave background radiation (CMBR). However, the early 
realizations of inflation require extremely flat potentials 
and very small coupling constants. To solve this naturalness 
problem, Linde has introduced \cite{Linde}, some years ago, the 
hybrid inflationary scenario which involves two real scalar 
fields, one of which may be a gauge non-singlet. Unfortunately, 
in this scheme, the (abrupt) termination of inflation is 
followed by a `waterfall' regime during which topological defects 
can be copiously produced \cite{smooth1}. In particular, the 
cosmological problem caused by the overproduction of GUT 
magnetic monopoles is not avoided in this inflationary scenario. 

\par
The simplest framework for realizing hybrid inflation is 
provided \cite{Cop,dvasha} by the supersymmetric (SUSY) GUTs
which are based on gauge groups with rank greater than five. 
The same superpotential which breaks spontaneously the GUT 
gauge group to a subgroup with lower rank also leads 
\cite{dvasha} to successful hybrid inflation with `natural' 
values of the relevant coupling constant. The slowly rolling 
inflaton field belongs to a gauge singlet superfield which 
couples to a conjugate pair of gauge non-singlet Higgs 
superfields. These fields acquire non-vanishing vacuum 
expectation values (vevs) after the end of inflation, thereby 
breaking the GUT gauge symmetry. The tree-level scalar 
potential possesses a flat valley of local minima for values 
of the gauge singlet inflaton greater than a certain critical 
value. The vevs of the Higgs superfields vanish along this 
valley. Inflation takes place when the system is trapped in 
the valley of local minima. The (classical) flatness of the 
valley is lifted by the one-loop radiative corrections 
\cite{dvasha} to the scalar potential. They are non-zero 
because of the SUSY breaking caused by the non-vanishing 
vacuum energy density on the valley. The slow-roll conditions 
(see e.g., Ref.\cite{cosmology}) are satisfied for all values
of the gauge singlet inflaton greater than its critical value, 
where inflation ends abruptly. It is followed by a `waterfall' 
regime during which the Higgs fields quickly fall towards the 
SUSY minima of the potential and oscillate about them. If the 
SUSY vacuum manifold is homotopically non-trivial, topological 
defects will be copiously formed by the Kibble mechanism since 
the system can end up at any point of the vacuum manifold with 
equal probability. So a cosmological disaster is encountered in 
the hybrid inflationary models which are based on GUTs 
predicting the existence of magnetic monopoles. 

\par
The monopole problem can be solved by utilizing the following 
observation \cite{pati}. The standard superpotential of SUSY 
hybrid inflation involves only renormalizable terms. Note, 
though, that an infinite number of non-renormalizable terms, 
which are linear in the gauge singlet inflaton, cannot be 
excluded by any symmetries. These terms are usually neglected. 
However, the leading of these terms, if its dimensionless 
coefficient is of order unity, can compete with the trilinear 
term of the standard superpotential whose coupling constant is 
typically of order $10^{-3}$. A `shifted' classically flat 
valley of local minima where the GUT gauge symmetry is broken 
then also appears \cite{pati} and can be used as an 
alternative inflationary trajectory. The necessary inclination 
along this valley is again given by the one-loop radiative 
corrections to the potential which are now calculated 
\cite{pati} with both the gauge singlet inflaton and the Higgs 
superfields acquiring constant non-zero values. This scheme is 
known \cite{lisbon} as shifted hybrid inflationary scenario. 
Inflation again ends abruptly by a `waterfall' with the system 
falling into the SUSY vacua. The main difference from standard 
(SUSY) hybrid inflation is that now no topological defects can 
form at the end of inflation since the GUT gauge symmetry is 
already broken during inflation. 

\par
In Ref.\cite{pati}, a concrete SUSY GUT model based on the 
Pati-Salam (PS) gauge group 
$G_{PS}=SU(4)_c\times SU(2)_L\times SU(2)_R$ and leading 
to successful shifted hybrid inflation has been constructed. 
The SUSY PS model is particularly interesting since it can also 
be obtained \cite{string} from string compactifications. It is 
one of the simplest GUT models predicting the existence of 
magnetic monopoles. These PS monopoles, which carry two units 
of `Dirac' magnetic charge \cite{magg}, would then lead to a 
cosmological disaster if standard hybrid inflation was employed. 
The results of the cosmic background explorer (COBE) 
\cite{cobe} were reproduced in this model with `natural' 
values of the relevant parameters and $G_{PS}$ spontaneous 
breaking scale of the order of $10^{16}~{\rm GeV}$. Moreover, 
the $\mu$ term was generated via a Peccei-Quinn (PQ) symmetry 
and proton was practically stable. Hierarchical light neutrino 
masses were generated via the seesaw mechanism. Baryogenesis in 
the universe could occur via a primordial leptogenesis 
\cite{lepto} consistently with the gravitino limit 
\cite{khlopov} on the `reheat' temperature, the $SU(4)_c$ 
symmetry and the experimental data on solar and atmospheric 
neutrino oscillations. 

\par
An alternative solution to the monopole problem of (SUSY) 
hybrid inflation had been given some years ago in 
Ref.\cite{smooth1}. The idea was to impose an extra discrete 
$Z_2$ symmetry which forbids the (renormalizable) trilinear 
coupling of the standard inflationary superpotential and 
introduce instead the leading non-renormalizable term. The
inflationary superpotential then coincides with the `shifted' 
one, but without the trilinear term. The emerging picture is, 
however, dramatically different from the one encountered in 
shifted hybrid inflation. The scalar potential possesses two 
symmetric valleys of local minima which are suitable for 
inflation and along which the GUT gauge symmetry is broken. 
The inclination of these valleys, which contain the SUSY minima 
too, is already non-zero at the classical level and thus 
the one-loop radiative corrections to the scalar potential are 
not needed. In Ref.\cite{smooth2}, it has been shown that, for 
almost all initial conditions, the system is trapped in either 
of these valleys and generates an adequate number of e-foldings 
as it moves towards the SUSY minimum following \cite{smooth1} 
the bottom of the valley. Inflation ends when the slow-roll 
conditions are violated and the system smoothly enters into an 
oscillatory phase about the SUSY minimum. This is the reason for 
calling this scenario smooth hybrid inflation. Since a specific 
inflationary path leading to a specific SUSY minimum is chosen 
already from the beginning of inflation, no topological defects 
can form at the end of inflation. In particular, the 
cosmological monopole problem is solved. The CMBR quadrupole 
anisotropy can be reproduced with `natural' values of the 
parameters and a gauge symmetry breaking scale identical to the 
SUSY GUT scale. 

\par
In this letter, we present a concrete SUSY GUT model which 
naturally leads to smooth hybrid inflation. It is based 
on the PS gauge group and is consistent with all the 
cosmological and phenomenological requirements. We pay 
particular attention to the study of the `reheating' process 
following inflation and the generation of the observed baryon 
asymmetry of the universe (BAU) via a primordial leptogenesis 
\cite{lepto} in this model.

\par
We start with the PS SUSY GUT model of Ref.\cite{pati}. The 
breaking of $G_{PS}$ down to the standard model gauge group 
is achieved by the superheavy vevs of the right-handed neutrino 
components of a conjugate pair of Higgs superfields 
\begin{eqnarray}
\bar{H}^c &=& (4,1,2) \equiv \left(\begin{array}{cccc}
                       \bar{u}^c_H & \bar{u}^c_H & 
                       \bar{u}^c_H &\bar{\nu}_H^c\\
                      \bar{d}^c_H & \bar{d}^c_H & 
                      \bar{d}^c_H & \bar{e}^c_H
                      \end{array}\right), \nonumber\\
H^c &=& (\bar{4},1,2) \equiv \left(\begin{array}{cccc}
                       u^c_H & u^c_H & u^c_H & \nu_H^c\\
                       d^c_H & d^c_H & d^c_H & e^c_H
                      \end{array}\right).
\end{eqnarray}
The three families of `matter' superfields are $F_i=(4,2,1)$ 
and $F^c_i=(\bar{4},1,2)$ ($i=1,2,3$), while the electroweak 
Higgs doublets $h^{(1)}$, $h^{(2)}$ belong to the superfield 
$h=(1,2,2)$. The model also contains a gauge singlet $S$ which 
triggers the breaking of $G_{PS}$, an $SU(4)_c$ 6-plet 
$G=(6,1,1)$ which gives masses to $\bar{d}^c_H$, $d^c_H$ 
(see first paper in Ref.\cite{string}), and a pair of gauge 
singlets $\bar{N}$, $N$ for solving \cite{rsym} the $\mu$ 
problem of the minimal SUSY standard model (MSSM) via a PQ 
symmetry which also solves the strong CP problem. In addition to 
$G_{PS}$, the model possesses two global $U(1)$ symmetries, 
namely a PQ and a R-symmetry, as well as a discrete $Z_2^{mp}$ 
symmetry (`matter 
parity') under which $F$, $F^c$ change sign. For details on the 
charge assignments and the full superpotential, the reader is 
referred to Ref.\cite{pati}. Baryon and lepton number are not 
conserved in this model and proton can decay via effective 
dimension five operators from one-loop graphs. However, its 
lifetime turns out to be long enough so that it is practically 
stable. The right-handed neutrinos $\nu_i^c$ acquire 
intermediate Majorana masses via non-renormalizable couplings 
to the superheavy vevs of $\bar{H}^c$, $H^c$. Light neutrino 
masses are then generated via the seesaw mechanism.

We impose \cite{smooth1} on the model an extra $Z_2$ symmetry 
under which $H^c\rightarrow-H^c$. The structure of the model 
remains unaltered except that now only even powers of the 
combination $\bar{H}^c H^c$ are allowed in the superpotential 
terms. In particular, all the properties summarized in the 
previous paragraph are still valid. The only physically 
significant alteration is that the trilinear term 
$S\bar{H}^c H^c$ is now missing from the inflationary 
superpotential. Including instead the leading non-renormalizable 
term, this superpotential becomes \cite{smooth1}
\begin{equation}
\delta W=S\left(-\mu^2+
\frac{(\bar{H}^c H^c)^2}{M_S^2}\right)\cdot
\label{eq:susyinfl}
\end{equation}
Here $\mu$ is a superheavy mass parameter and 
$M_S\sim 5\times 10^{17}~{\rm GeV}$ is the string mass scale. 
The dimensionless coupling constants have been absorbed in $\mu$ 
and $M_S$. Note that these two mass parameters can be made 
positive by field redefinitions.

\par
The inflationary scalar potential $V$ derived from $\delta W$ 
in Eq.(\ref{eq:susyinfl}) is given by \cite{smooth1}
\begin{equation}
\tilde{V}=\frac{V}{\mu^4}=(1-\tilde\chi^4)^2+
16\tilde\sigma^2\tilde\chi^6, 
\end{equation}
where we have used the dimensionless fields $\tilde\chi=\chi/2
(\mu M_S)^{1/2}$ and $\tilde\sigma=\sigma/2
(\mu M_S)^{1/2}$ with $\chi$, $\sigma$ being normalized real 
scalar fields defined by $\bar{\nu}_H^c=\nu_H^c=\chi/2$, 
$S=\sigma/\sqrt{2}$ after rotating $\bar{\nu}_H^c$, 
$\nu_H^c$, $S$ to the real axis by appropriate gauge and 
R-transformations. This potential has a completely different 
structure from both the standard and `shifted' hybrid 
inflationary potentials. It still possesses a flat direction at 
$\tilde\chi=0$, but this is now a local maximum with respect to 
$\tilde\chi$ for all values of $\tilde\sigma$. It also has 
\cite{smooth1} two symmetric valleys of local minima with 
respect to $\tilde\chi$ at
\begin{equation}
\tilde\chi=\pm\sqrt{6}\tilde\sigma\left[\left(1+
\frac{1}{36\tilde\sigma^4}\right)^{\frac{1}{2}}-1
\right]^{\frac{1}{2}},
\label{eq:smoothvalley}
\end{equation}  
which can be used as inflationary trajectories. They contain the 
SUSY vacua which lie at $\tilde\chi=\pm 1$, $\tilde\sigma=0$. 
Note that these valleys are not classically flat. In fact, 
already at the tree level, they possess an inclination, which 
can drive the inflaton towards the SUSY vacua. As a consequence,
contrary to the case of standard SUSY or shifted hybrid 
inflation, there is no need of radiative corrections, which are 
expected to give a subdominant contribution to the slope of the 
inflationary paths. In spite of this, one could try to include 
the one-loop corrections. This requires the construction of the 
mass spectrum on the inflationary trajectories. In doing so, we 
find that the ${\rm mass}^2$ of some scalars belonging to the 
inflaton sector is negative. The one-loop corrections, which 
involve logarithms of the masses squared, are then ill-defined. 
This may be remedied by resumming the perturbative expansion to 
all orders, which is a formidable task and we do not pursue it 
here. 

\par
The dimensionless potential along the inflationary valleys is 
given by
\begin{equation}
\tilde{V} = 48\tilde\sigma^4\left[72\tilde\sigma^4\left(
1+\frac{1}{36\tilde\sigma^4}\right)\left(\left(1+
\frac{1}{36\tilde\sigma^4}\right)^{\frac{1}{2}}-1\right)
-1\right]. 
\label{eq:smoothV}
\end{equation}
The system follows \cite{smooth1}, from the beginning, a 
particular inflationary path and, thus, ends up at a specific 
point of the vacuum manifold leading to no production of 
monopoles. Also, no disastrous domain walls from the 
spontaneous breaking of the extra $Z_2$ symmetry under which 
$H^c$ changes sign are generated. The termination of inflation 
is not abrupt as in the other two hybrid inflationary scenarios. 
The reason is that the inflationary path is stable with respect 
to $\tilde\chi$ for all values of $\tilde\sigma$. Inflation 
ends smoothly at a value $\tilde\sigma_0$ of $\tilde\sigma$ 
after which the slow-roll conditions cease to hold. The leading 
slow-roll parameters $\epsilon$ and $\eta$ (see e.g., 
Ref.\cite{cosmology}) involve the derivatives of the potential 
along the inflationary path, which are 
\begin{equation}
\frac{d\tilde{V}}{d\tilde\sigma}=192\tilde\sigma^3
\left[(1+144\tilde\sigma^4)\left(\left(1+
\frac{1}{36\tilde\sigma^4}\right)^{\frac{1}{2}}
-1\right)-2\right],
\label{eq:firstder}
\end{equation}
\begin{eqnarray}
\frac{d^2\tilde{V}}{d\tilde\sigma^2}&=&
\frac{16}{3\tilde\sigma^2}
\Biggl\{(1+504\tilde\sigma^4)
\left[72\tilde\sigma^4\left(\left(1+
\frac{1}{36\tilde\sigma^4}\right)^{\frac{1}{2}}
-1\right)-1\right]
\nonumber \\
& &-(1+252\tilde\sigma^4)\left(\left(1+
\frac{1}{36\tilde\sigma^4}\right)^{-\frac{1}{2}}
-1\right)\Biggl\}.
\label{eq:secondder}
\end{eqnarray}

\par
We calculate the quadrupole anisotropy of the CMBR and the 
number of e-foldings, $N_Q$, of our present horizon during 
inflation using the standard formulae (see e.g., 
Ref.\cite{cosmology}) and Eq.(\ref{eq:firstder}). The 
present inflationary scenario has an important advantage. 
The common vev of $\bar{H}^c$ and $H^c$ at the SUSY vacuum, 
which equals $(\mu M_S)^{1/2}$, is not so tightly restricted 
as in the previous hybrid inflationary scenarios. Using this 
freedom, we choose it equal to the SUSY GUT scale 
$M_G\approx 2.86\times 10^{16}~{\rm GeV}$. From the 
results of COBE \cite{cobe} and for $N_Q\approx 57$, we 
then obtain $M_S\approx 4.39\times 10^{17}~{\rm GeV}$ and 
$\mu\approx 1.86\times 10^{15}~{\rm GeV}$, which are quite
`natural'. The value of $\sigma$ at which inflation ends 
corresponds to $\eta=-1$ and is $\sigma_0\approx 1.34\times 
10^{17}~{\rm GeV}$ ($\epsilon$ remains always much smaller 
than unity). Finally, the value of $\sigma$ at which our 
present horizon crosses outside the inflationary horizon is  
$\sigma_Q \approx 2.71 \times 10^{17}~{\rm GeV}$.

\par
We now turn to the discussion of the `reheating' process 
following inflation and the generation of the observed BAU. 
After the end of inflation, the system smoothly enters into a 
phase of damped oscillations about the SUSY vacuum. The 
oscillating inflaton fields are two complex scalars $\theta=
(\delta\bar\nu^c_H+\delta\nu^c_H)/\sqrt{2}$ 
($\delta\bar\nu^c_H$, $\delta\nu^c_H$ are the deviations 
of $\bar\nu^c_H$, $\nu^c_H$ from their common vev) and $S$. 
They have the same mass $m_{\rm infl}=2\sqrt{2}\mu^2/M_G
\approx 3.42\times 10^{14}~{\rm GeV}$ and decay into 
right-handed neutrinos and sneutrinos respectively via the 
superpotential terms 
$\gamma_i\bar{H}^c\bar{H}^cF^c_iF^c_i/M_S$ (in a basis 
where the $\gamma$'s are diagonal) and 
$S(\bar{H}^cH^c)^2/M_S^2$. Their common decay width is 
\begin{equation}
\Gamma=\frac{1}{8\pi}
\left(\frac{M_i}{M_G}\right)^2m_{\rm infl}~,
\label{eq:gamma}
\end{equation} 
where $M_i=2\gamma_i\mu$ is the mass of the heaviest 
$\nu_i^c$ satisfying the inequality $M_i<m_{\rm infl}/2$. To 
minimize the number of small coupling constants, we assume that
\begin{equation}
M_2<\frac{1}{2}m_{\rm infl}\leq M_3=2\gamma_3 \mu~,
\label{eq:ineq}
\end{equation}
so that the oscillating inflaton fields decay into the second 
heaviest right-handed neutrino superfield $\nu^c_2$. For MSSM 
spectrum, the `reheat' temperature $T_{r}\approx 
(1/7)(\Gamma M_P)^{1/2}=(1/7)(M_2/M_G) 
(m_{\rm infl}M_P/8\pi)^{1/2}$ \cite{lss}, 
where $M_P\approx 1.22\times 10^{19}~{\rm GeV}$ is the 
Planck mass scale. Hence, $M_2$ can be expressed in terms of 
$T_r$. The subsequent decay of $\nu^c_2$ into lepton and 
electroweak Higgs superfields creates a primordial lepton 
asymmetry \cite{lepto} which survives \cite{ibanez} until 
the electroweak transition, where it is converted in part into 
baryon asymmetry via non-perturbative electroweak sphaleron 
effects. 

\par
Analysis \cite{giunti} of the CHOOZ experiment \cite{chooz} 
shows that the solar and atmospheric neutrino oscillations 
decouple, allowing us to concentrate on the two heaviest 
families. The light neutrino mass matrix, which is generated 
via the seesaw mechanism, is
\begin{equation}
m_{\nu}\approx -\tilde{m}^{D}\frac{1}{M^R}m^{D}, 
\label{eq:neut1}
\end{equation}
where $m^D$ is the `Dirac' neutrino mass matrix and $M^R$ the
Majorana mass matrix of right-handed neutrinos with (positive) 
eigenvalues $m^D_{2,3}$ and $M_{2,3}$ respectively. The two 
(positive) eigenvalues of $m_{\nu}$ are denoted by 
$m_2=m_{\nu_{\mu}}$ and $m_3=m_{\nu_{\tau}}$. 
The determinant and trace invariance of $m_\nu^\dag m_\nu$ 
provide us with two constraints \cite{Laz3} on the mass 
parameters $m_{2,3}$ , $m^D_{2,3}$ , $M_{2,3}$ and the 
rotation angle $\theta$ and phase $\delta$ which diagonalize 
$M^R$ in the basis where $m^D$ is diagonal. The primordial 
lepton asymmetry is given \cite{Laz3} by
\begin{equation}
\frac{n_L}{s}\approx 1.33\ \frac{9T_r}{16\pi 
m_{\rm infl}}\frac{M_2}{M_3}\ \frac{c^2s^2
\sin 2\delta (m_3^{D\hspace{0.05cm}^2}-
m_2^{D\hspace{0.05cm}^2})^2}{|\langle h^{(1)}\rangle|^2
(m_3^{D\hspace{0.05cm}^2}s^2+
m_2^{D\hspace{0.05cm}^2} c^2)}~,
\label{eq:nls}
\end{equation}
where $c=\cos\theta$, $s=\sin\theta$ and the electroweak 
vev coupling to the up-type quarks 
$|\langle h^{(1)}\rangle|\approx 174~\rm{GeV}$ since, 
here, $\tan\beta$ is large as implied \cite{pana} by the 
fact that the electroweak Higgs as well as the right-handed 
quark superfields form $SU(2)_R$ doublets. For MSSM spectrum, 
$n_L/s\approx -(79/28)(n_B/s)$, where $n_B/s$ is the 
BAU (see second paper in Ref.\cite{ibanez}). Note that 
Eq.(\ref{eq:nls}) holds provided \cite{pilaftsis} that 
$M_2 \ll M_3$ and the decay width of $\nu_3^c$ is 
$\ll (M_3^2-M_2^2)/M_2$, which are well-satisfied here. The 
$\mu-\tau$ mixing angle $\theta_{23}=\theta_{\mu\tau}$ 
lies \cite{Laz3} in the range
\begin{equation}
|\varphi-\theta^D|\leq\theta_{\mu\tau}\leq
\varphi+\theta^D~,~~{\rm for}~~\varphi+\theta^D\leq\pi/2~,
\label{eq:mixing}
\end{equation}
where $\varphi$ is the rotation angle which diagonalizes 
$m_\nu$ in the basis where $m^D$ is diagonal and $\theta^D$ 
is the `Dirac' mixing angle defined with vanishing Majorana 
masses for the $\nu^c$'s.

\par
We are now ready to examine whether the gravitino constraint 
\cite{khlopov} on the `reheat' temperature and the restrictions 
on the BAU from Big Bang nucleosynthesis can be satisfied 
consistently with the data on solar and atmospheric neutrino 
oscillations and the $SU(4)_c$ symmetry. The gravitino 
constraint is usually quoted as 
$T_r\stackrel{_{<}}{_{\sim }}10^9~{\rm GeV}$. 
Unfortunately, we do not find any solutions for such values of 
$T_r$. Therefore, we take $T_r=10^{10}~{\rm GeV}$, which is 
also perfectly acceptable provided \cite{kawasaki} that the 
branching ratio of the gravitino to photons is somewhat smaller 
than unity and the gravitino mass is relatively large (of order 
a few hundred GeV). The low deuterium abundance constraint 
\cite{deuterium} on the BAU, 
$0.017\stackrel{_{<}}{_{\sim }}\Omega_Bh^2
\stackrel{_{<}}{_{\sim }}0.021$, yields the bound 
$1.8\times 10^{-10}\stackrel{_{<}}{_{\sim }}-n_L/s
\stackrel{_{<}}{_{\sim }}2.3\times 10^{-10}$. The small 
or large mixing angle MSW solution of the solar neutrino 
puzzle requires 
$2\times 10^{-3}~{\rm eV}\stackrel{_{<}}{_{\sim }}
m_{\nu_\mu}\stackrel{_{<}}{_{\sim }} 3.2\times 
10^{-3}~{\rm eV}$ or $3.6\times 10^{-3}~{\rm eV}
\stackrel{_{<}}{_{\sim }}m_{\nu_\mu}
\stackrel{_{<}}{_{\sim }} 13\times 10^{-3}~{\rm eV}$ 
respectively \cite{bahcall}. The $\tau$-neutrino mass is 
restricted in the range $3\times 10^{-2}~{\rm eV}
\stackrel{_{<}}{_{\sim }}m_{\nu_\tau}
\stackrel{_{<}}{_{\sim }}9\times 10^{-2}~{\rm eV}$ 
from the results of the SuperKamiokande experiment \cite{osc} 
which also imply almost maximal $\nu_\mu-\nu_\tau$ mixing, 
i.e., $\sin^2 2\theta_{\mu\tau}
\stackrel{_{>}}{_{\sim }}0.85$. We assume, for simplicity, 
that the `Dirac' mixing angle $\theta^D$ is negligible, so that 
$\theta_{\mu\tau}\approx\varphi$. The $SU(4)_c$ symmetry 
implies that $m^D_3$ coincides asymptotically with the top 
quark mass. Taking renormalization effects into account, for 
MSSM spectrum with large $\tan\beta$, we obtain \cite{Laz3} 
$m^D_3\approx 110-120~{\rm GeV}$. We also include the running 
of $\theta_{\mu\tau}$ from $M_G$ to the electroweak scale 
(see last paper in Ref.\cite{lepto}). 

\par
\begin{figure}[ht]
\psfig{figure=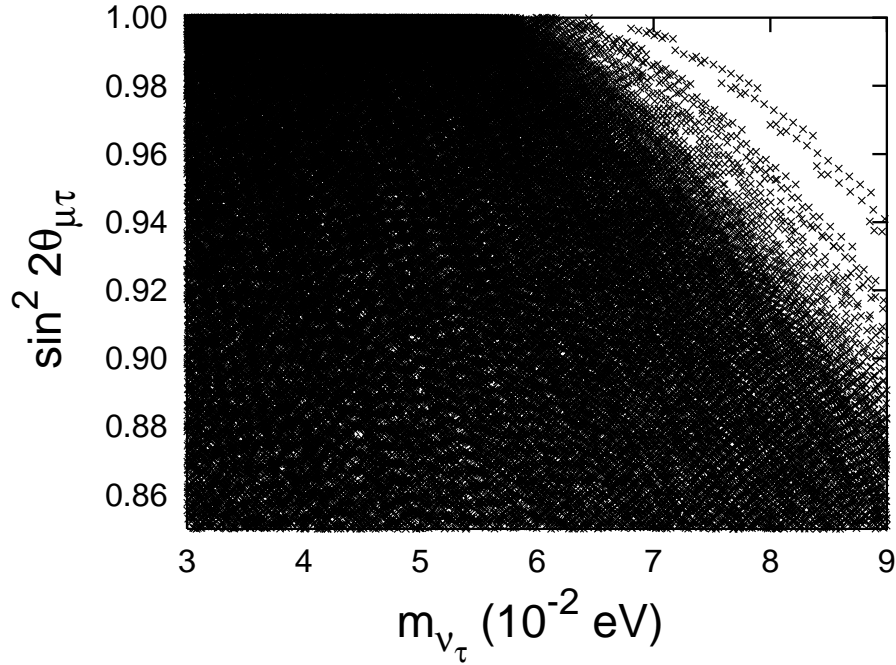,height=9cm,width=13.96cm}
\caption{The scatter plot in the $m_{\nu_\tau}-
\sin^2 2 \theta_{\mu\tau}$ plane of the solutions which 
satisfy the low deuterium abundance constraint on the BAU, the 
restrictions from solar and atmospheric neutrino oscillations, 
and the $SU(4)_c$ invariance. We take $T_r=10^{10}~{\rm GeV}$, 
$\gamma_3\approx 0.05-0.5$, $m^D_2\approx 0.8-2~{\rm GeV}$ 
and $m^D_3\approx 110-120~{\rm GeV}$.}
\label{sin2n1}
\end{figure}
\begin{figure}[ht]
\psfig{figure=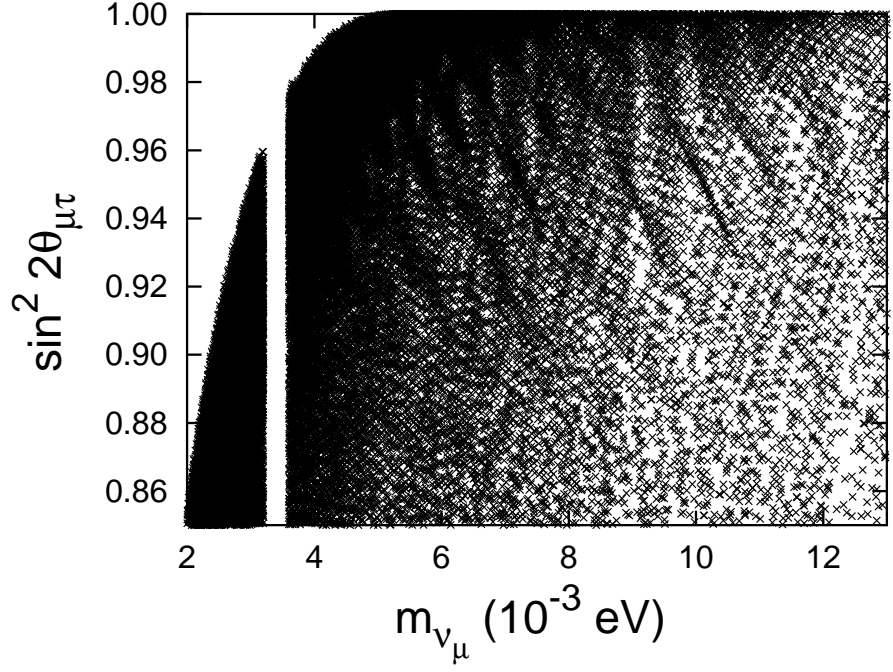,height=9cm,width=13.96cm}
\caption{The scatter plot in the $m_{\nu_\mu}-
\sin^2 2\theta_{\mu\tau}$ plane of the solutions depicted in 
Fig.\ref{sin2n1}.}
\label{sin2n2}
\vskip 0.5cm
\end{figure}
\begin{figure}[ht]
\psfig{figure=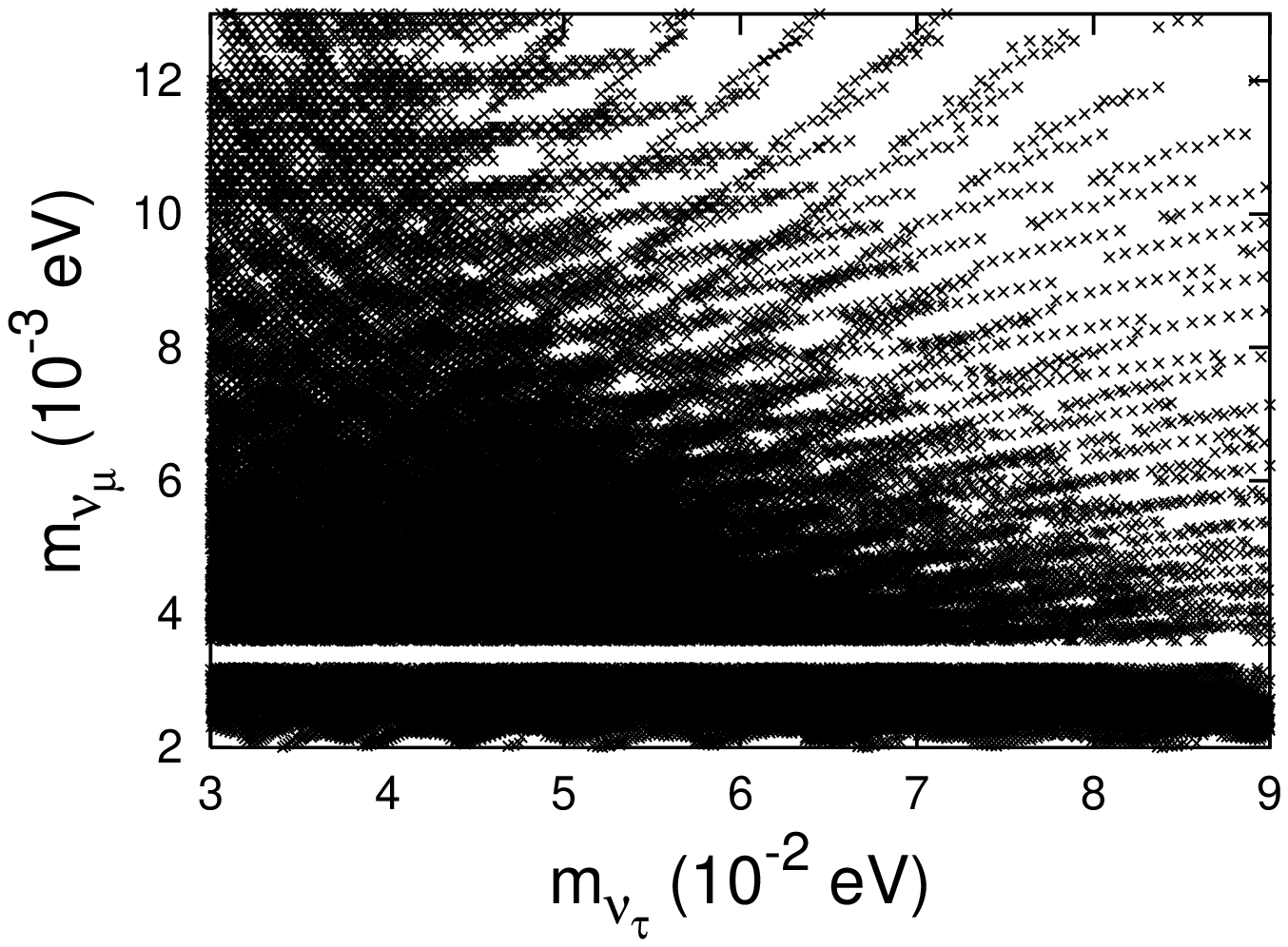,height=9cm,width=13.96cm} 
\caption{The scatter plot in the $m_{\nu_\tau}-m_{\nu_\mu}$ 
plane of the solutions depicted in Fig.\ref{sin2n1}.}
\label{n12}
\vskip 0.5cm
\end{figure}

Our results are shown in Figs.\ref{sin2n1}, \ref{sin2n2} and 
\ref{n12}, where we plot solutions corresponding to 
$T_r=10^{10}~{\rm GeV}$ and satisfying the leptogenesis 
constraint consistently with the neutrino oscillation data and 
the $SU(4)_c$ symmetry. The parameter $\gamma_3$ runs from 
$0.05$ to $0.5$, i.e., $M_3\approx 1.86\times 10^{14}-1.86
\times 10^{15}~{\rm GeV}$. The second inequality in 
Eq.(\ref{eq:ineq}) implies that 
$\gamma_3\stackrel{_{>}}{_{\sim }}0.046$. However, no 
solutions are found for $\gamma_3<0.05$. Also, values of 
$\gamma_3$ higher than 0.5 do not allow solutions. The mass 
of the second heaviest right-handed neutrino 
$M_2\approx 1.55\times 10^{11}~{\rm GeV}$, which 
corresponds to $T_r=10^{10}~{\rm GeV}$ and clearly 
satisfies the first inequality in Eq.(\ref{eq:ineq}). The 
restrictions from $SU(4)_c$ invariance are expected to be 
more or less accurate only if applied to the masses of the 
third family quarks and leptons. For the second family, they 
should hold only as order of magnitude relations. We thus 
restrict ourselves to values of $m^D_2$ smaller than 
$2~{\rm GeV}$ since much bigger $m^D_2$'s would violate 
strongly the $SU(4)_c$ symmetry (the value of $m^D_2$ from 
exact $SU(4)_c$ is \cite{Laz3} about $0.23~{\rm GeV}$ for 
MSSM spectrum with large $\tan\beta$). Moreover, we find 
that solutions exist only if 
$m^D_2\stackrel{_{>}}{_{\sim }}0.8$. So we take 
$m^D_2\approx 0.8-2~{\rm GeV}$ and, 
as required by $SU(4)_c$ invariance, $m^D_3\approx 110-120
~{\rm GeV}$. Also, the phase 
$\delta\approx (-\pi/8)-(-\pi/5)$ and the rotation angle 
$\theta\approx 0.01-0.03$ for solutions to appear. Note that 
$\delta$'s close to 0 or $-\pi/2$ are excluded since they 
yield very small primordial lepton asymmetry. 

\par
In Fig.\ref{sin2n1}, we present the scatter plot of our 
solutions in the $m_{\nu_\tau}-\sin^2 2\theta_{\mu\tau}$ 
plane. In Figs.\ref{sin2n2} and \ref{n12}, solutions are 
plotted in the $m_{\nu_\mu}-\sin^2 2\theta_{\mu\tau}$ 
and $m_{\nu_\tau}-m_{\nu_\mu}$ planes respectively. Our 
model favors the large mixing angle MSW solution to the solar 
neutrino problem. For large $m_{\nu_\tau}$'s (close to the 
upper bound) or small $m_{\nu_\mu}$'s (close to the lower 
bound), it excludes very large $\mu-\tau$ mixing and, in any 
case, disfavors large $m_{\nu_\tau}$'s. 

\par
In conclusion, we presented a concrete PS SUSY GUT model leading 
naturally to smooth hybrid inflation, which solves the 
cosmological magnetic monopole problem. The model reproduces the 
measured quadrupole anisotropy of the CMBR with `natural' values 
of the parameters and a PS spontaneous breaking scale equal to 
the SUSY GUT scale. A PQ symmetry is used to generate the $\mu$ 
term of MSSM and proton is practically stable. Inflation is 
followed by a successful `reheating' process satisfying the 
gravitino constraint on the `reheat' temperature and generating 
the observed BAU via a primordial leptogenesis consistently with 
the requirements from solar and atmospheric neutrino oscillations 
and the $SU(4)_c$ symmetry. 
\vspace{0.5cm}

\par
This work was supported by European Union under the TMR contract 
ERBFMRX-CT96-0090 and the RTN contracts HPRN-CT-2000-00148 and 
HPRN-CT-2000-00152. One of us (S. K.) was supported by PPARC.

\def\ijmp#1#2#3{{ Int. Jour. Mod. Phys. }{\bf #1~}(#2)~#3}
\def\ijmpa#1#2#3{{ Int. Jour. Mod. Phys. }{\bf A#1~}(#2)~#3}
\def\pl#1#2#3{{ Phys. Lett. }{\bf B#1~}(#2)~#3}
\def\zp#1#2#3{{ Z. Phys. }{\bf C#1~}(#2)~#3}
\def\prl#1#2#3{{ Phys. Rev. Lett. }{\bf #1~}(#2)~#3}
\def\rmp#1#2#3{{ Rev. Mod. Phys. }{\bf #1~}(#2)~#3}
\def\prep#1#2#3{{ Phys. Rep. }{\bf #1~}(#2)~#3}
\def\pr#1#2#3{{ Phys. Rev. }{\bf D#1~}(#2)~#3}
\def\np#1#2#3{{ Nucl. Phys. }{\bf B#1~}(#2)~#3}
\def\mpl#1#2#3{{ Mod. Phys. Lett. }{\bf #1~}(#2)~#3}
\def\arnps#1#2#3{{ Annu. Rev. Nucl. Part. Sci. }{\bf
#1~}(#2)~#3}
\def\sjnp#1#2#3{{ Sov. J. Nucl. Phys. }{\bf #1~}(#2)~#3}
\def\jetp#1#2#3{{ JETP Lett. }{\bf #1~}(#2)~#3}
\def\app#1#2#3{{ Acta Phys. Polon. }{\bf #1~}(#2)~#3}
\def\rnc#1#2#3{{ Riv. Nuovo Cim. }{\bf #1~}(#2)~#3}
\def\ap#1#2#3{{ Ann. Phys. }{\bf #1~}(#2)~#3}
\def\ptp#1#2#3{{ Prog. Theor. Phys. }{\bf #1~}(#2)~#3}
\def\plb#1#2#3{{ Phys. Lett. }{\bf#1B~}(#2)~#3}
\def\apjl#1#2#3{{ Astrophys. J. Lett. }{\bf #1~}(#2)~#3}
\def\n#1#2#3{{ Nature }{\bf #1~}(#2)~#3}
\def\apj#1#2#3{{ Astrophys. Journal }{\bf #1~}(#2)~#3}
\def\anj#1#2#3{{ Astron. J. }{\bf #1~}(#2)~#3}
\def\mnras#1#2#3{{ MNRAS }{\bf #1~}(#2)~#3}
\def\grg#1#2#3{{ Gen. Rel. Grav. }{\bf #1~}(#2)~#3}
\def\s#1#2#3{{ Science }{\bf #1~}(#2)~#3}
\def\baas#1#2#3{{ Bull. Am. Astron. Soc. }{\bf #1~}(#2)~#3}
\def\ibid#1#2#3{{ ibid. }{\bf #1~}(#2)~#3}
\def\JHEP#1#2#3{{ JHEP }{\bf #1~}(#2)~#3}
\def\npps#1#2#3{{ Nucl. Phys. (Proc. Sup.) }{\bf B#1~}(#2)~#3}
\def\astp#1#2#3{{ Astropart. Phys. }{\bf #1~}(#2)~#3}
\def\epj#1#2#3{{ Eur. Phys. J. }{\bf C#1~}(#2)~#3}
\def\stmp#1#2#3{{ Sprin. Trac. Mod. Phys. }{\bf #1~}(#2)~#3}

\vspace{0.5cm}

\end{document}